\documentstyle[12pt]{article}
\begin{document}
\pagestyle{empty}
\setlength{\baselineskip}{0.1875in}
\newcommand{\psection}[1]{\vspace{\baselineskip}
\noindent {\bf #1} \vspace{\baselineskip}}
\def\thefootnote{\fnsymbol{footnote}}

%%definitions
\newcommand{\nc}{\newcommand}
\nc{\beq}{\begin{equation}}   \nc{\eeq}{\end{equation}}
\nc{\beqa}{\begin{eqnarray}}  \nc{\eeqa}{\end{eqnarray}}
\nc{\lsim}{\begin{array}{c}\,\sim\vspace{-21pt}\\< \end{array}}
\nc{\gsim}{\begin{array}{c}\sim\vspace{-21pt}\\> \end{array}}

{\hbox to\hsize{ \hfill March 1994}}
\begin{flushleft}
{\bf IN SEARCH OF CLASSICAL TRAJECTORIES
\footnotemark[1]}
\end{flushleft}
\vspace{\baselineskip}
\begin{tabbing}
\hspace{1.0in} \= Thomas M. Gould \\[\baselineskip]
\> Department of Physics and Astronomy\\
\> The Johns Hopkins University\\
\> Baltimore, MD 21218
\end{tabbing}
\footnotetext[1]{To appear in the  Proceedings of the 1994 NATO Advanced
Research Workshop,Sintra, Portugal}

\vspace{\baselineskip}

\psection{INTRODUCTION}

In this short talk,
I will emphasize some of the key points of a new Minkowski space
formulation of nonperturbative contributions to quantum
scattering~\cite{GHP},
described at length by Hsu elsewhere in this volume.
I will especially comment on some of the practical considerations
of computing classical trajectories.

Our formulation~\cite{GHP} is based on a stationary phase
approximation to a scattering amplitude involving an initial
two-particle state.
A real or complex classsical stationary trajectory is determined
by an initial value or boundary value problem, respectively.
The work establishes  the role of Minkowski trajectories for
quantum scattering amplitudes in weakly coupled field theories,
and outlines a procedure for finding such trajectories,
in principle via computation.
Phenomena of particular interest in this regard involve final
states with
(i) large numbers of particles ($N\sim 1/g^2$ where
$g$ is the coupling constant of the theory),
and/or (ii) anomalous violation of global quantum numbers,
like fermion number violation in the electroweak theory.

Standard perturbative methods cannot describe these phenomenon.
Current nonperturbative Euclidean space methods
break down at high energies ($E\sim M_w/g^2$ in electroweak theory)
because they do not account for the non-vacuum boundary conditions
appropriate for a scattering problem.
Our approach is to find the stationary point of a scattering amplitude,
to include the effects of initial and final states on the stationary
trajectory,
and then to assess in what regime the expansion around this
classical trajectory is controlled.

The formalism was illustrated for the simple case of a real scalar
field~\cite{GHP}.
A distinction is then made between real and complex stationary trajectories
of the scattering amplitude of the real scalar field.
I emphasize this here because the computation of the two types of trajectories
could in practice be quite different.

\setlength{\topskip}{0.0in}
\vspace{\baselineskip}
\psection{REAL TRAJECTORIES}

A given initial state on a time slice $t=T_i$
implies initial values for $\phi$ and $\dot{\phi}$,
which are sufficient to determine a real trajectory
uniquely~\cite{GHP}.
The initial condition expressed in terms of Fourier components of
the field is
\beq
\label{phialpha}
\phi_i(\vec{k}) ~=~ {1\over \sqrt{2\,\omega_{\bf k}}}~
\left(\,
u_{\bf k}~e^{-i\omega_k T_i} ~+~ u_{-\bf k}^*~e^{i\omega_k T_i}
\,\right) ~,
\eeq
where the negative frequency component is
\beq
\label{u}
u_{\bf k} ~=~
{
\alpha_R(\vec{k} ) ~+~ \alpha_L(\vec{k}) \over
\left(\,
1 ~+~ \int d^3k ~\alpha_R(\vec{k} )~\alpha_L(\vec{k})
\,\right)^{1/2}
}~,
\eeq
with the (normalized) right- and left-moving  wave packets
$\alpha_R(\vec{k} )$ and $\alpha_L(\vec{k})$ of the two-particle
initial state.
One assumes here that the field behaves freely for times
$t<T_i$, so that the field and its time derivative are related in
the obvious way.
The field evolves forward in time from (\ref{phialpha})
according to the source-free equations of motion
\beq
\label{eom}
{\delta S \over \delta \phi(x)} ~=~ 0 ~.
\eeq

So, a real trajectory is determined by an initial value problem.
A real trajectory connects every initial wave packet state with a
unique final state.
The corresponding scattering amplitude is then not exponentially
suppressed, since its action is real.
But, the final state may or may not be
interesting from the point of view of (i) and (ii) above.
So, one must search the space of fields for interesting trajectories,
by varying the initial wave packets.
One might hope to find evidence of instabilities for the generation
of long wavelength modes from the initial short wavelength modes,
the signal of many (soft) particles in the final state.

I stress that
a negative result to the search does not rule out the existence of
unsuppressed amplitudes involving interesting final states.
There is no one-to-one correspondence between real trajectories and
scattering amplitudes.  (more on this below)
There may well be amplitudes not dominated
in any way by real classical trajectories !

\psection{Hints for Classical Trajectories}

Several previous investigations into the classical behavior of
a variety of field theories provide some hints about
the behavior of their real trajectories.
I refer you to the discussion and references contained in Gould et
al.~\cite{GHP}
Here I would like to mention one most promising situation: evidence
for an instability in Yang-Mills theory.
M\"uller et al.~\cite{GMMT}
have considered the classical stability of a stationary
mono-color wave in Yang-Mills theory
\mbox{(in $A^c_0 = 0$ gauge):}
\beq
\label{stwv}
A^c_i (x,t) ~=~ \delta_{i3} ~\delta_{c3} ~A ~\cos k_0x ~\cos \omega_0 t,
\eeq
where $c$ is a color and $i$ is a spatial index.
Small amplitude variations of the field in directions of different
($c\neq 3$) color are found to lead to an instability with long wavelength.
This implies the existence of a classical trajectory connecting initial
high energy plane waves to long wavelength modes in the final state.
It therefore suggests that the corresponding $2 \rightarrow$ {\it many}
gluon scattering amplitude may be unsuppressed !

For the purpose of the scattering problem,
it would be interesting to see whether the instability in Yang-Mills
theory persists for wave packets of finite spatial extent.
As the width of a wave packet in momentum space decreases,
approaching the plane wave limit (\ref{stwv})
in which the instability has been observed,
its amplitude decreases also, with the normalization held fixed~\cite{GHP}.
However, since the instability persists for arbitrarily small
amplitude~\cite{GMMT},
one may hope that it still produces many long wavelength modes.

In the context of electroweak theory, it will also be necessary
to consider how the instability is modified in the presence
of an explicit symmetry breaking scale.
Some benchmark figures for a study of electroweak theory
might be the following.
One could consider initial wave packets with average momenta
$$
k_{avg} \sim M_w/g^2 ~,
$$
and width
$$
\Delta k \sim M_w ~.
$$
Such wave packets overlap for a very short time for weak coupling,
$\sim g^2/M_w$,
during which an  instability must generate long
wavelength amplitudes.
I emphasize though that a quantitative understanding of these estimates
should be obtainable from direct computation of classical trajectories
in a chosen field theory.

\vspace{\baselineskip}
\psection{COMPLEX TRAJECTORIES}

Complex trajectories of the real field arise from an analytic
continuation of the path integral describing the scattering
amplitude~\cite{GHP}.
Complex trajectories can in principle connect
a given initial state with {\em any} final state,
due to the additional degrees of freedom.
They include the case of classically forbidden transitions,
in the case where initial and final states are based on different
vacua separated by an energy barrier in configuration space.
This is the situation relevant for processes with anomalous
fermion number violation in electroweak theory and center of mass
energies below the sphaleron energy, $\sim M_w/g^2$.
Complex trajectories will have complex actions,
so the corresponding amplitudes can in general have some
exponential suppression.

The field is initially free, so it has a form
similar to (\ref{phialpha})
\beq
\label{phicomplex}
\phi_i(\vec{p}) ~=~ {1\over \sqrt{2\omega_{\bf p}}}~
\left(\,
u_{\bf p}~
e^{-i\omega_p T_i} ~+~  v_{\bf p}~e^{i\omega_p T_i}
\,\right) ~,
\eeq
but where $u_{\bf p}$ and $v_{\bf p}$ are now independent complex functions.
The initial boundary condition  is satisfied by
\beq
\label{ucomplex}
u_{\bf p} ~=~
{\alpha_R(\vec{p})\over\int d^3k~\alpha_R(\vec{k})~v_{\bf k}}~+~
{\alpha_L(\vec{p})\over\int d^3k~\alpha_L(\vec{k})~v_{\bf k}}~,
\eeq
where $v_{\bf k}$ is arbitrary.
The initial  configurations which satisfy
(\ref{phicomplex}), (\ref{ucomplex}) have negative frequency modes
which ``resemble'' the wave packet $\alpha_L + \alpha_R$ (up to
complex multiplicative factors) and positive frequency modes which
are arbitrary.

An additional degree of freedom is present, which must be determined
by the additional information in the final state.
The final state enforces a condition on the positive frequency part
of the complex trajectory at final time $T_f$.
The negative frequency part of the field remains independent and undetermined.
The boundary condition takes the form
\beq
\label{phicomplexf}
\phi_f(\vec{k}) ~=~ {1\over \sqrt{2\,\omega_{\bf k}}}~
\left(\,
c_{\bf k}~
e^{-i\omega_k T_f} ~+~  b_{-\bf k}^*~e^{i\omega_k T_f}
\,\right) ~,
\eeq
for a final state $\vert {\bf b}^* \rangle$.
Finally, the field evolves again with the source-free equations of motion,
(\ref{eom}).

So, a complex trajectory is determined by a boundary value problem,
with boundary conditions given by the initial and final states.
One can consider solving this problem with a search procedure similar
to the case of a real trajectory.  A given $u_{\bf p}$ specified by
the initial state gives rise to any number of final states, by
varying $v_{\bf k}$.
Alternatively, a direct solution of the boundary value problem for
given initial and final states may be possible.

\vspace{\baselineskip}
\psection{A WINDOW OF INTEREST}

Once an interesting classical trajectory is found,
one must turn to the problem of the quantum corrections
to determine the regime
in which the semiclassical expansion is controlled.
I would like to emphasize in this regard how our semiclassical expansion
differs subtly from the more familiar instanton expansion.
In the familiar semiclassical expansion around an instanton,
the vacuum boundary conditions of the classical field
allow us to scale the field by the coupling constant,
both in the boundary conditions and in the action
$$
S\left[ \phi \right] = {1\over g^2}~\hat{S}\left[ g\phi\right]
$$
so that  $\hat{S}$ is independent of the coupling $g$.
Then, the expansion is good, and the instanton contribution dominates,
for sufficiently small coupling $g$, where the semiclassical exponent
is large.

In our formulation,
we account for the non-vacuum boundary conditions
relevant to a scattering amplitude.
As such, the boundary conditions are fixed by physical initial and/or final
states, and the scaling above cannot be made.
Now,
we must require the coupling to be sufficiently large to produce
an instability for the production of long wavelength modes from
short wavelength modes.
Meanwhile, our experience with ordinary perturbation theory and
the semiclassical expansion leads us to expect that the
the coupling must be sufficiently  small to control the expansion.
Thus, there must be a {\em window} in the coupling in order that
our method be both interesting and controlled.
While we have not proven that a window exists in any field theory,
the means by which to do so are clear.
The upper bound on the coupling can be determined by computations
of classical solutions to the field equations, of the type outlined
above.
The lower bound on the coupling will have to come from computations
of the quantum corrections, as discussed in Gould et al.~\cite{GHP}

\vspace{\baselineskip}
\psection{COMMENTS}

Current methods are inadequate to describe nonperturbative contributions
to scattering amplitudes at very high energies.
An outstanding example is the unsolved problem of the rate of
fermion number violation in high energy collisions ($E\sim M_w/g^2$)
in electroweak theory.
I will close by emphasizing that {\em any} solution of the classical
problems which I have outlined here corresponds to the stationary
point of {\em some} scattering amplitude,
and therefore yields {\em some} nontrivial information about the
nonlinear aspects of quantum field theory,
entirely inaccessible to perturbation theory.
While much is known about semiclassical calculations in the vacuum sectors
of field theories,
our work points towards the need for a study of a much wider class
of classical solutions, involving multi-particle boundary conditions.

\psection{Acknowledgements}

The author acknowledges the support of the National Science Foundation
under grant \mbox{NSF-PHY-90-96198}.

%Special abbreviated journal names for Sintra proceedings
%Will convert old bibitems to Sintra format automatically.
\nc{\pr}[3]{        {\em Phys. Rev.  }#1:#3 (19#2) }

\vspace{\baselineskip}

\end{document}